\documentstyle[12pt]{article}
\def\beqar {\begin{eqnarray}}
\def\eeqar {\end{eqnarray}}
\def\beq {\begin{equation}}
\def\eeq {\end{equation}}
\def\A{{\cal A}}

\def\C{{\cal C}}

\def\J{{\cal J}}

\def\P{{\cal P}}

\def\N{{\cal N}}
\def\M{{\cal M}}

\def\Om{\Omega}

\def\si{\sigma}
\def\Si{\Sigma}

\def\p{\phi}
\def\d{\partial}

\def\Ad{{\dot A}}

\def\<{\langle}\def\bra{\langle}
\def\>{\rangle}\def\ket{\rangle}
\def\ka{\left(\frac{\kappa}{2}\right)}
\def\Tr{{\rm Tr}}

\def\cp{{\bf CP}}
\begin{document}

\begin{titlepage}
\null\vspace{-62pt} \pagestyle{empty}
\begin{center}
\vspace{1.0truein}

{\Large\bf An interpretation of multigraviton amplitudes} \\

\vspace{1.0in} YASUHIRO ABE \\
\vskip .1in {\it Physics Department\\ City College of the
CUNY\\
New York, NY 10031}\\
\vskip .05in {\rm E-mail: abe@sci.ccny.cuny.edu}\\
\vspace{1.5in}
\centerline{\large\bf Abstract}
\end{center}
We obtain alternative expressions for the multigraviton tree level
amplitudes and discuss their general properties. In particular, by
analogy with Yang-Mills theory, we find that some combinatoric structure
can be carried by a Chan-Paton factor of general relativity as a gauge
theory.

\end{titlepage}
\pagestyle{plain} \setcounter{page}{2} 


Recently, there has been remarkable progress in the computation
and understanding of multigluon amplitudes based on Witten's
formulation of twistor string theory \cite{Witten1}. What is
proposed in the twistor string theory is an equivalence between
the perturbative $\N=4$ super Yang-Mills theory and the
D-instanton expansion of a topological string theory, the
so-called topological B-model of the supertwistor space
$\cp^{3|4}$. This equivalence implies that any sorts of multigluon
tree level amplitudes can be obtained from the topological B-model
of $\cp^{3|4}$. An important example of such a correspondence was
in fact observed a number of years ago by Nair \cite{Nair1} in the
cases of the so-called maximally helicity violating (MHV)
amplitudes \cite{PT}. It was pointed out that the multigluon MHV
amplitudes could be obtained as current correlators of a
Wess-Zumino-Witten (WZW) model with a natural interpretation in
superwistor space. A wide variety of calculations for the
multigluon amplitudes based on the twistor string proposal has
been carried out and shows complete agreement with known results,
providing remarkable simplification of computing processes from
the standard field theoretic techniques. For those amplitudes
whose expressions have not been known otherwise, twistor string
theory has also been used to obtain new results. Progress along
these lines including loop calculations can be found in
\cite{Mason}.

One of the significant features in these recent calculations of
amplitudes is the use of two-component spinors in parametrizing
the external momenta of scattering gluons (or massless particles),
with the spinors being identified to the homogeneous coordinates
on $\cp^1$. As is well-known \cite{PT}, the use of spinor momenta
facilitates the helicity based computation of the amplitudes. In
addition, the $\cp^1$ on which the spinors are defined is crucial
to interpret the multigluon amplitudes in a framework of the
twistor space. What is dramatic in twistor string theory is that
one can obtain the helicity based amplitudes by relating the
$\cp^1$ of spinors to a certain algebraic curve in twistor space;
for example, the tree level MHV amplitude corresponds to a
degree-one curve (or a straight line), the tree level next-to-MHV
amplitude to a degree-two curve and, etc. Since the spinor
momentum can be used for massless particles of any spins, it is
natural to ask for an interpretation of multigraviton amplitudes
in the same twisor oriented framework. In fact, the gravitational
MHV amplitudes were considered from this point of view in
\cite{Witten1} and it was seen that the MHV amplitudes for general
relativity could also be related to the degree-one curves in
twistor space. (For expositions of other cases, see
\cite{S-Brook}.) Lately, there has been much attention to the
calculation of multigraviton amplitudes along these lines. Some of
the latest reports can be found in \cite{Bern3,Nair2,QM,CS,Bohr1}.
(For a review of multigraviton amplitudes in general, one can
refer to \cite{Bern1}.)

Certainly, the twistor space is useful in an analysis of massless
particles, however, direct use of twistor space for the
calculations of multigraviton amplitudes seems to have problems as
we discuss in the following. Twistor space $\cp^3$ can be
considered as a $\cp^1$ bundle over compactified spacetime ${\bf
R}^4$ (or $S^4$) \cite{Penrose}. The four-dimensional local
spacetime coordinates defined as such preserves conformal
invariance. Any gravitational theory which emerges from twistor
string theory is expected to naturally preserve conformal
invariance. It is important to understand conformal supergravity
in twistor stirng theory especially in connection with the loop
calculations of multigluon amplitudes. This was first considered
intensively in \cite{BW1}. (For related investigations, see
\cite{Ahn}.) Conformal supergravity is, however, not something we
would like to consider in an analysis of multigraviton amplitudes,
since it is not a well-defined theory; unitarity is believed to be
broken and hence there is no known S-matrix of the theory. (For a
review of conformal supergravity, see \cite{csg}.) This is
cumbersome because, contrary to the case of multigluon amplitudes,
the lack of S-matrix prevents us from making practical use of
twistor string theory in physically reasonable gravitational
models.

Unlike conformal gravity, Einstein gravity (or general relativity)
does not have such a problem and one can in principle calculate
the amplitudes with it. In fact, it is known that the
multigraviton tree level amplitudes can be expressed in terms of
the multigluon counterparts by use of the Kawai-Lewellen-Tye (KLT)
relation between tree amplitudes of closed and open string
theories \cite{KLT}. Berends, Giele and Kuijf applied this KLT
relation to the four-dimensional spacetime by taking the field
theory limit ($\alpha' \rightarrow 0$) and showed a general way of
expressing the multigraviton tree level amplitudes in terms of the
multigluon ones \cite{BGK}. In particular, the multigraviton MHV
amplitudes were explicitly obtained in terms of the spinor momenta
of scattering gravitons. It turns out that, unlike the Yang-Mills
cases, the gravitational MHV amplitudes are highly nonholomorphic
in terms of the spinor momenta. Recently, these multigraviton MHV
amplitudes were analyzed in \cite{Nair2} and it was suggested that
$\N=8$ supergravity could emerge from some version of Berkovits'
alternative twistor string theory. Note that in the alternative
twistor string theory it is proposed that the perturbative $\N=4$
Yang-Mills theory is also equivalent to the perturbative expansion
of a certain open string theory, rather than the D-instanton
expansion of the topological string theory \cite{Berk1}. The
recent suggestion implies that, at least for the MHV cases, it is
possible to obtain the multigraviton tree level amplitudes from
open string tree amplitudes with a suitable definition of its
Chan-Paton factor. The above mentioned nonholomorphic factors are
interpreted as the Chan-Paton factor of such an open string
amplitude.

The correspondence between the graviton MHV amplitudes and the
open string theory sounds novel because the latter does not
usually contain gravitons. This implies a duality between gauge
theory and gravity theory at weak coupling, rather than the
well-known strong-weak duality of the two theories \cite{Malda}.
Generalization of this (weak-weak) correspondence to the non-MHV
amplitudes is exactly what we attempt to pursue in what follows.
We shall consider a general form of the multigraviton tree level
amplitudes inspired by the KLT relation. Rather than resorting to
the spinor momenta, we will simply express the multigraviton
amplitudes in terms of the corresponding multigluon amplitudes. We
then show that an open-string (or a gauge-theory) interpretation
of the multigraviton amplitudes is also applicable to the non-MHV
amplitudes in general, with their Chan-Paton factors being
understood as suggested in \cite{Nair2}.


Let us begin with a brief review of the KLT-inspired amplitude. An
explicit expression of the tree level scattering amplitude for $n$
gravitons is given by Bern {\it et al} \cite{Bern2} in the
following form  \beqar \M(12\cdots n) &=& i \ka^{n-2}
(-1)^{n+1}\sum_{\si \in S_{r-1}} \sum_{\tau \in S_{r-2}} f(\si)
\tilde{f}(\tau) \nonumber\\&& ~~~~\times \, \C(12\cdots n) ~
\C(\si_2 \cdots \si_r \,1 \, n-1 \,
\tau_{r+1} \cdots \tau_{n-2} \,n) \nonumber\\
&&~+ ~\P(23\cdots n-2) \label{am1}\eeqar where $\kappa=\sqrt{32\pi
G_N}$, with $G_N$ being Newton's constant, and $\P(23 \cdots n-2)$
indicates the terms obtained by the permutations of the elements
$\{2,3, \cdots, n-2 \}$. The summations are taken over $\si$ and
$\tau$ which denote the permutations of $\{2,3,\cdots, r\}$ and
$\{r+1, \cdots, n+2 \}$ respectively,
{\it i.e.}, $\si=\left(%
\begin{array}{c}
  2 \cdots r \\
  \si_2 \cdots \si_r \\
\end{array}%
\right)$, $\tau=\left(%
\begin{array}{c}
  r+1 \cdots n-2 \\
  \tau_{r+1} \cdots \tau_{n-2} \\
\end{array}%
\right)$. Note that the expression (\ref{am1}) corresponds to the
amplitudes for the even number of gravitons, $n=2r$. A similar
expression can be obtained for odd $n \ge 3$, as we will see in a
moment. The factor $\C(12\cdots n)$ comes from the tree level
scattering amplitude for $n$ gluons given by \beq {\cal A}(12
\cdots n) = i g^{n-2} \sum_{\si \in S_{n-1}}\Tr (t^{a_1}
t^{a_{\si_2}}\cdots t^{a_{\si_n}}) ~\C(1 \si_2 \cdots \si_n)
\label{am2} \eeq where $g$ is the Yang-Mills coupling constant and
$t^a$'s are the generators of the gauge group $U(N)$ in the
fundamental representation. The summation here is taken over the
permutations of the elements $\{2,3, \cdots, n \}$. Notice that
the momentum-conservation $\delta$-functions are implicit both in
(\ref{am1}) and (\ref{am2}). The factor $f(\si) \tilde{f}(\tau)$
in (\ref{am1}) is defined by \beqar f(\si)= f(\si_2 \cdots \si_r)
&=& \{1\si_r\} \prod_{m=2}^{r-1} \left[
\{1 \si_m \} + \sum_{k=m+1}^{r} g(\si_k \, \si_m) \right] \label{am3}\\
\tilde{f}(\tau)= \tilde{f}(\tau_{r+1} \cdots \tau_{n-2}) &=&
\{\tau_{r+1} \,n-1\} \prod_{m=r+2}^{n-2} \left[ \{\tau_m \, n-1
\} + \sum_{k=r+1}^{m-1} g(\tau_m \,\tau_k) \right] \label{am4}\\
g(i\, j) &=&  \left\{
    \begin{array}{ll}
      \{i \, j\} & \mbox{for $i<j$}
\\
      0  & \mbox{otherwise}
    \end{array} \right. \label{am5} \eeqar
where the curly bracket denotes a product of external momenta
carried by gravitons, {\it i.e.}, $\{i \, j\} = \{p_i \, p_j\} =
p_i \cdot p_j$. Conventionally, this product is defined in terms
of two component indices, {\it i.e.}, $p_i \cdot p_j= (p_i)_{A\Ad}
(p_j)^{A\Ad}$ where $A = 1,2$, $\Ad = 1,2$.

For odd $n$, the amplitudes $\M(12\cdots n)$ is obtained by
replacing the permutation $S_{r-2}$ with $S_{r-1}$ in (\ref{am1}).
For example, the amplitude with $n=7$, $r=3$ contains the factor
$\tilde{f}(\tau_4 \tau_5)$, instead of just $\tilde{f}(\tau_4)$ as
in the case with $n=6$, $r=3$.


So far, we simply recapitulate a known result. Let us consider a
different way of writing this result. Generically, scattering
amplitudes can be expressed as a number of functional derivatives
acting on an S-matrix functional. This naturally explains the term
$\P(23\cdots n-2)$ in $\M(12\cdots n)$ as well as $\P(23 \cdots
n)$ in $\A(12\cdots n)$. The sums over $\si \in S_{r-1}$ and $\tau
\in S_{r-2}$ in $\M$ arise from an internal structure in this
context. One can then ask why the sum is taken in such a way that
splits the elements $\{2,3,\cdots, n-2\}$ in the middle. The
question brings us back to the original computation of the KLT
relation \cite{KLT}, where the sum has been chosen such that the
number of terms in $\M$ decreases most efficiently. The sum can in
fact be taken in an arbitrary way in terms of the separation of
the elements $\{2,3,\cdots, n-2\}$. The factor $f(\si)
\tilde{f}(\tau)$ remains calculable from (\ref{am3})-(\ref{am5}),
with $r=n/2$ being replaced by some other number, say $n-3$. Note
that if we shift $r$, the number of terms in $\M$ increases
accordingly.

One might further ask if there is a more homogeneous way of taking
the sum, since the amplitudes should preserve the bosonic symmetry
in transpositions of the gravitons. As we will see shortly, it
turns out that the amplitude (\ref{am1}) can indeed be rewritten,
provided that we introduce a particular sum over the elements
$\{i_1, i_2, \cdots, i_r, j_1, j_2, \cdots, j_{n-r} \} = \{1, 2,
\cdots, n \}$ such that the relations, $i_1 < i_2 < \cdots < i_r$
and $j_1 < j_2 < \cdots < j_{n-r}$, are preserved. One can think
of this sum as `homogeneous' because, unlike the elements $\si$'s
and $\tau$'s in (\ref{am1}), $i$'s and $j$'s are now evenly
defined as the elements of $\{ 1,2, \cdots, n \}$. Let us consider
rewriting the amplitude (\ref{am1}) by use of this `homogeneous'
sum. For simplicity, we first look at the case of $n=6$. Writing
down every term in (\ref{am1}), we find that the amplitude can
also be expressed as \beqar \hspace{-1.2cm} \M(12\cdots 6) &=& -i
\ka^{4} \sum_{\{\si_2,\si_3,\tau_4
\}=\{2,3, 4 \}} ~  [ ~  \C(1 \si_2 \si_3 \tau_4 5 6 ) \nonumber \\
&& \hspace{-0.6cm}\times \left( ~ f'(\si_2 \si_3)
\tilde{f}'(\tau_4) \, \C(\si_2 \si_3  1 5  \tau_{4}  6)~ +
~\P(\si_2 \si_3 |  \tau_{4})~ \right)~ + ~\P(\si_2 \si_3  |
\tau_{4}) ~ ] \label{lap3} \eeqar where the sum is taken over
$\{\si_2, \si_3, \tau_4\}=\{2,3,4\}$ such that $\si_2 < \si_3$,
namely, there are only three cases, $(\si_2 \si_3 \tau_4)=(234),
(243),$ and $(342)$. Since the elements $\{ \si_2, \si_3, \tau_4
\}$ are now defined in a homogeneous way, the factor $f'(\si_2
\si_3) \tilde{f}'(\tau_4)$ is slightly different from $f(\si_2
\si_3) \tilde{f}(\tau_4)$. Explicit forms of $f'$ and $\tilde{f}'$
are given by \beqar f'(23) = \{12\}\{13\} ~,~ f'(32)=
\{12\}\{(1+2)3\} ~,~ \tilde{f}'(4)=\{45\}\nonumber\\
f'(24) = \{12\}\{14\} ~,~ f'(42) =
\{12\}\{(1+2)4\} ~,~ \tilde{f}'(3)=\{35\}\label{lap4}\\
f'(34) = \{13\}\{14\} ~,~ f'(43) = \{13\}\{(1+3)4\} ~,~
\tilde{f}'(2)=\{25\}\nonumber \eeqar Notice that $f'$ and
$\tilde{f}'$ are the same as $f$ and $\tilde{f}$ when $(\si_2
\si_3 \tau_4)=(234)$, and the rest of the terms in (\ref{lap4})
can be obtained by the replacements $(234) \leftrightarrow (243)$
and $(234) \leftrightarrow (342)$. In this sense, the factor
$f'(\si_2 \si_3) \tilde{f}'(\tau_4)$ is essentially the same as
$f(\si_2 \si_3) \tilde{f}(\tau_4)$. The symbol $\P(\si_2 \si_3 |
\tau_{4})$ in (\ref{lap3}) denotes the permutations of $\si$ and
$\tau$. Since there is only one entry for $\tau$, $\P(\si_2 \si_3
| \tau_{4})$ here means simply $\si_2 \leftrightarrow \si_3$. In
(\ref{lap3}), the first permutation inside the parenthesis should
be taken in cooperation with the expressions in (\ref{lap4}),
while the second permutation at the end is simply applied to the
terms obtained as such. Generalization of the amplitude
(\ref{lap3}) can be done in the following form \beqar \M(12\cdots
n) &=& i \ka^{n-2} (-1)^{n+1} \sum_{\{\si,\tau \}=\{2,3, \cdots,
n-2 \}} ~ [ ~ \C(1 \si_2 \cdots \si_r \tau_{r+1}
\cdots \tau_{n-2} \, n-1 \, n ) \nonumber \\
&& \hspace{-1.6cm}\times \left( ~ f'(\si) \tilde{f}'(\tau) \,
\C(\si_2 \cdots \si_r \, 1 \, n-1 \, \tau_{r+1} \cdots \tau_{n-2}
\, n) + \P(\si_2 \cdots \si_r
|  \tau_{r+1} \cdots \tau_{n-2})~ \right) \nonumber\\
&& \hspace{6.5cm} + \P(\si_2 \cdots \si_r  |  \tau_{r+1} \cdots
\tau_{n-2}) ~ ] \label{lap5} \eeqar Notice that the number of
terms in the amplitude remains the same as $(r-1)! (r-2)! (n-3)!$
($n=2r$). What we have basically done is to rewrite the factor
$(n-3)!$ as $ {}_{n-3} C_{r-1} (r-1)! (r-2)!$, where ${}_{n-3}
C_{r-1} = \frac{(n-3)!}{(r-1)!(r-2)!}$ corresponds to the number
of $\{\si,\tau \}$-combinations involving the `homogeneous' sum.
(Since the sum preserves the orderings $\si_2 < \cdots < \si_r$
and $\tau_{r+1} < \cdots < \tau_{n-2}$, one can easily find that
the number of possible combinations for the elements $\{\si_2,
\cdots \si_r, \tau_{r+1}, \cdots, \tau_{n-2} \}$ is ${}_{n-3}
C_{r-1}$.) The missing $(r-1)!(r-2)!$ factor arises from the the
second permutation $\P(\si_2 \cdots \si_r | \tau_{r+1} \cdots
\tau_{n-2})$ which is equivalent to $\P(\si_2 \cdots \si_r) \times
\P( \tau_{r+1} \cdots \tau_{n-2})$. The double appearance of
$\P(\si_2 \cdots \si_r  | \tau_{r+1} \cdots \tau_{n-2})$ in
(\ref{lap5}) implies that the functional derivative for gravitons
can be represented by a composite derivative when $\M$ is to be
expressed in a functional language.

As we have seen in (\ref{lap4}), the factors $f'(\si)\tilde{f}'(\tau)$
in (\ref{lap5}) can essentially be obtained from $f(\si)
\tilde{f}(\tau)$ via (\ref{am3})-(\ref{am5}). By shifting the parameter
$r$, we can in fact further simplify the computation of these factors,
as we see in the following. Let us consider the case of $n=6$ again. The
amplitude (\ref{lap3}) can be rewritten as \beqar \M(12\cdots 6) &=& -i
\ka^{4} \left[~ \C(1 2 3 4 5 6 ) \left(~ f(23) \{45\} ~\C(23 1 5 4 6) +
2 \leftrightarrow 3~ \right) +
2 \leftrightarrow 3~ \right] \nonumber\\
&& ~~~~ +
\left(%
\begin{array}{c}
  2 \\
  3 \\
  4 \\
\end{array}%
\right) \leftrightarrow \left(%
\begin{array}{c}
  2 \\
  4 \\
  3 \\
\end{array}%
\right)  +
\left(%
\begin{array}{c}
  2 \\
  3 \\
  4 \\
\end{array}%
\right) \leftrightarrow \left(%
\begin{array}{c}
  3 \\
  4 \\
  2 \\
\end{array}%
\right) \label{lap6}\eeqar Notice that all terms in $\M(12\cdots 6)$ can
be deduced from the knowledge of $f(23) = \{12\}\{13\}$ and $f(32) =
\{12\}\{(1+2)3\}$. Shifting $r$ from $n/2$ to $n-3$ in (\ref{lap5})
(which is possible as we have discussed earlier), we can generalize the
expression (\ref{lap6}) in the following form
 \beqar \M(12\cdots n) &=&
i \ka^{n-2} (-1)^{n+1} ~~[~ \C(1 2 \cdots n) ~~(~ f(23\cdots n-3)~ \{n-2
\, n-1 \} \nonumber\\&& ~~~\times ~ \C(23 \cdots n-3 \,1 \, n-1 \, n-2
\, n) + \P(23\cdots n-3) ~ )
 + \P(23\cdots n-3) ~ ] \nonumber\\
&& \hspace{-2.4cm} +
\left(%
\begin{array}{c}
  2 \\
  3 \\
  \vdots \\
  n-4 \\
  n-3 \\
  n-2 \\
\end{array}%
\right) \leftrightarrow \left(%
\begin{array}{c}
  2 \\
  3 \\
  \vdots \\
  n-4 \\
  n-2 \\
  n-3 \\
\end{array}%
\right)  +
\left(%
\begin{array}{c}
  2 \\
  \vdots \\
  n-5 \\
  n-4 \\
  n-3 \\
  n-2 \\
\end{array}%
\right) \leftrightarrow \left(%
\begin{array}{c}
  2 \\
  \vdots \\
  n-5 \\
  n-3 \\
  n-2 \\
  n-4 \\
\end{array}%
\right) + \cdots +
\left(%
\begin{array}{c}
  2 \\
  3 \\
  \vdots \\
  n-4 \\
  n-3 \\
  n-2 \\
\end{array}%
\right) \leftrightarrow \left(%
\begin{array}{c}
  3 \\
  4 \\
  \vdots \\
  n-3 \\
  n-2 \\
  2 \\
\end{array}%
\right) \nonumber\\ \label{lap7}\eeqar which can be expressed
alternatively as \beqar \M(12\cdots n) &=& i \ka^{n-2} (-1)^{n+1}
\sum_{\{\si_2 < \si_3 < \cdots < \si_{n-3}\, , \, \tau_{n-2}\}} ~[~ \C(1
\si_2 \cdots\si_{n-3} \tau_{n-2}\, n-1 \, n) \nonumber\\&&
\hspace{-1.9cm} \times \left(~ f'(\si_2 \si_3 \cdots \si_{n-3})~
\{\tau_{n-2} \, n-1 \}  ~ \C(\si_2 \si_3 \cdots \si_{n-3}\, 1 \, n-1 \,
\tau_{n-2} \, n) + \P(\si_2 \si_3 \cdots \si_{n-3}) ~ \right) \nonumber\\
&& \hspace{6.8cm} ~~~~~~~~~~~
 + \P(\si_2 \si_3\cdots \si_{n-3}) ~ ] \label{lap8}\eeqar where the sum
is taken over the elements $\{\si_2 , \si_3 , \cdots , \si_{n-3},
\tau_{n-2}\} = \{2, 3, \cdots , n-2\}$ such that it preserves the
ordering $\si_2 < \si_3 < \cdots < \si_{n-3}$. These amplitudes are
applicable for any number of gravitons (which is more than three). The
number of terms in $\M(12\cdots n)$ now becomes $(n-4)!(n-3)!$, rather
than the minimum value $\left(\frac{n-2}{2}\right)!
\left(\frac{n-4}{2}\right)!(n-3)!$ (for even $n$) corresponding to the
KLT-inspired form in (\ref{am1}).


As we have seen in (\ref{lap7}), the multigraviton amplitudes can
fully be expressed in terms of $f(23\cdots n-3)$ plus the
permutations of the elements $\{2,3,\cdots, n-3\}$. Such factors
can explicitly be obtained from the formula (\ref{am3}) as \beqar
f(\si_2 \si_3 \cdots \si_{n-3}) &=& \{1 \si_{n-3}\}
\prod_{m=2}^{n-4} \{\, (1+ \sum_{k=m+1}^{n-3}
\si_{k<m})\, \si_m \} \label{b1}\\
\si_{i<j} &=&  \left\{
    \begin{array}{ll}
      \si_i & \mbox{for $\si_i < \si_j$}
\\
      0  & \mbox{otherwise}
    \end{array} \right. \label{b2} \eeqar
Notice that (\ref{b1}) can be expressed as a product of \beqar
T_{\si_r} &\equiv& \{\, (1+\si_{r+1 < r} + \si_{r+2 <r} +\cdots
+\si_{n-3 < r}) \, \si_r \,\} ~~~~~\mbox{for $r=2,3,\cdots, n-4$}
\label{b3}\\ T_{\si_{n-3}} &\equiv& \{1\,\si_{n-3} \} \label{b4}
\eeqar For example, in the case of $n=7$, explicit forms of
$f(\si_2 \si_3 \si_4)$ are given by
\beqar f(234) &=& \{12\}\{13\}\{14\} \nonumber\\
f(243) &=& \{12\}\{13\}\{(1+3)4\} \nonumber\\ f(324)
&=&\{12\}\{(1+2)3\}\{14\} \nonumber\\f(342) &=&
\{12\}\{(1+2)3\}\{(1+2)4\} \label{b5}\\f(423) &=&
\{12\}\{13\}\{(1+2+3)4\} \nonumber\\f(432) &=&
\{12\}\{(1+2)3\}\{(1+2+3)4\}\nonumber \eeqar


Let us consider the graviton amplitude $\M$ of the form in
(\ref{lap8}). Following the above discussion, we can express the
factor $f'(\si_2 \si_3 \cdots \si_{n-3})$ and $\{\tau_{n-2}\,
n-1\}$ in (\ref{lap8}) as \beqar f'(\si_2 \si_3 \cdots \si_{n-3})
 &=& T_{\si_2} ~ T_{\si_3} ~ \cdots ~
T_{\si_{n-3}}  \label{g1} \\  \{\tau_{n-2}\, n-1\} &\equiv&
T_{\tau_{n-2}} \label{g1-a} \eeqar where $T_{\si_2}, T_{\si_3},
\cdots, T_{\si_{n-3}}$ are the same as (\ref{b3}), (\ref{b4})
except that the elements $\{\si_2 \si_3 \cdots \si_{n-3}\}$ are
now defined in $\{2,3, \cdots n-2\}$ instead of in $\{2,3,\cdots,
n-3\}$. The $\si$'s are exactly determined once the element
$\tau_{n-2}$ is picked out of $\{2,3,\cdots, n-2\}$ and the
concrete forms of $T_{\si_2}, T_{\si_3}, \cdots, T_{\si_{n-3}}$
can be straightforwardly obtained from (\ref{b3}), (\ref{b4}) by
replacing the elements $\{2,3,\cdots, n-4, n-3\}$ with
$\{2,3,\cdots, n-4, n-2\}$, $\{2,3,\cdots, n-5, n-3, n-2\}$, and
so on, as shown in the last line of (\ref{lap7}). In what follows,
we implicitly consider the subamplitudes of $\M(12\cdots n)$ in
which the element $\tau_{n-2}$ is fixed.

Let us remind that the factor $\C(12\cdots n)$ in $\M$ can
essentially be obtained from the corresponding gluon amplitude
$\A(12 \cdots n)$ in (\ref{am2}). It is known that, at least in
the case of the maximally helicity violating (MHV) amplitudes, we
can express the amplitude as \cite{Nair1}  \beqar \A(12\cdots n)
&\sim& \Tr (t^{a_1} t^{a_{2}}\cdots t^{a_{n}}) ~\C(1 2 \cdots n)
~+ ~\P(2 3 \cdots n) \nonumber\\ &\sim& \Tr (t^{a_1} t^{a_2}\cdots
t^{a_n}) ~ \bra A_{1}^{a_1} A_{2}^{a_2} \cdots A_{n}^{a_n} \ket
~+~ \P(2 3 \cdots n) \nonumber\\ &\sim& ~\bra A_{1} A_{2} \cdots
A_{n} \ket ~+~ \P(2 3 \cdots n) \label{g2}\eeqar where $A_m =
t^{a_m} A_{m}^{a_m}$ ($m=1,2, \cdots, n$) and $\bra A_1 A_2 \cdots
A_n \ket$ can be interpreted as a current correlator of a suitably
defined Wess-Zumino-Witten (WZW) model. In \cite{Nair1}, it was
realized that this WZW model had to contain $\N=4$ supersymmetry
in order to obtain the correct structure of the MHV amplitude.
(Later, we will further discuss this point in connection with
twistor space.) The computation of the correlator can efficiently
be carried out by introducing free fermions to represent the
current operators. With this representation, the trace or
Chan-Paton factor in $\A$ naturally arises from a series of
contractions among the neighboring free fermions. In the graviton
amplitude $\M$, there is no trace factor. In order to express the
MHV version of $\C(12\cdots n)$ purely, we need to introduce a
series of fixed-ordered WZW currents, $J_1 J_2 \cdots J_n$, and
its expectation value evaluated in a certain state, say, the
vacuum, on which the free fermion operators act, {\it i.e.},
$\C(12\cdots n) \sim \<0|J_{1} J_{2} \cdots J_{n}|0\>$. (For the
realization of such a vacuum state, one can refer to
\cite{Nair2}.) The current $J_m$ corresponds to the abelian (or
non-matrix) part of $A_m$ or $A_{m}^{a_m}$. Notice that
correlators of such currents can not be used to describe
$\C(12\cdots n)$, since without the trace factor there are
arbitrary ways of contracting fermions, which leads to an
inconsistency in defining the correlators of Abelian current. For
the moment, we will continue to consider the MHV case. The general
cases are also expected to satisfy the following discussion as we
will explain shortly.

In (\ref{lap8}), there are $\C(1 \si_2 \cdots \si_{n-3} \tau_{n-2}
n-1 \, n)$ and $\C(\si_2 \si_3 \cdots \si_{n-3} 1 \, n-1\,
\tau_{n-2} n)$. We assign $J$'s for the elements of the first $\C$
and another independent set of $J$'s for the elements of the
second $\C$. We will denote the other set by $\tilde{J}$,
satisfying also $\C(12\cdots n) \sim \< 0| \tilde{J}_1 \tilde{J}_2
\cdots \tilde{J}_n |0 \>$. The ingredients of the two $\C$'s along
with the factor $T_{\si_2} T_{\si_3} \cdots T_{\si_{n-3}}$ in
(\ref{g1}) can be listed by the following sequence \beqar &&
\hspace{-1.2cm}(T_{\si_2} \tilde{J}_{\si_2}) J_1 ~ (T_{\si_3}
\tilde{J}_{\si_3}) J_{\si_2} ~ \cdots ~(T_{\si_{n-3}}
\tilde{J}_{\si_{n-3}})
J_{\si_{n-4}} ~(T_{1} \tilde{J}_{1}) J_{\si_{n-3}} \nonumber\\
&& \hspace{3.8cm} ~~~\times ~(T_{n-1} \tilde{J}_{n-1})
J_{\tau_{n-2}}~(T_{\tau_{n-2}} \tilde{J}_{\tau_{n-2}}) J_{n-1}
~(T_{n} \tilde{J}_{n}) J_{n} \label{g3}\eeqar where $J$'s and
$\tilde{J}$'s follow the orderings of the first and second $\C$'s,
respectively. Since $T_\si$'s do not have any ordering issue, we
can make them couple to $\tilde{J}$'s as indicated. We also
include the couplings involving the factors
$T_{1}=T_{n-1}=T_{n}=1$ and $T_{\tau_{n-2}} = \{ \tau_{n-2}
n-1\}$. Let us denote $\J_1 =(T_{\si_2} \tilde{J}_{\si_2}) J_1 $,
$\J_{\si_2}=(T_{\si_3} \tilde{J}_{\si_3}) J_{\si_2}$ and so on, up
to $\J_n = (T_n \tilde{J}_n) J_n$. The expression of $\J_1$
implies that $J_1$ couples to $(T_{\si_2} \tilde{J}_{\si_2})$. We
choose this coupling simply because the element $1$ of the first
$\C$ is located in the same place as the element $\si_2$ of the
second $\C$. Since the ordering information on $\{\si_2, \cdots,
\si_{n-3}, \}$ is already encoded by $T_\si$'s for fixed
$\tau_{n-2}$, we can potentially choose any $\tilde{J}_\si$'s to
be coupled with $J_1$. The same analysis holds for the rest of
$J$'s. Thus, it would be natural to express an expectation value
of $\J_1 \J_2 \cdots \J_n$ in the following form \beqar \bra 0|
\J_1 \J_2 \cdots \J_n |0 \ket &=& \bra 0| ~ \left[ (T_{\si_2}
\tilde{J}_{\si_2}) (T_{\si_3} \tilde{J}_{\si_3}) \cdots (T_{n}
\tilde{J}_{n}) \right]_{perm} J_1 J_{\si_2} \cdots J_{n} ~|0 \ket
\nonumber\\ &=& \left[~ T_{\si_2}T_{\si_{3}}\cdots T_{n} ~ \bra 0|
\tilde{J}_{\si_2} \tilde{J}_{\si_3} \cdots \tilde{J}_n |0 \ket ~+~
\P(\si_2 \si_3 \cdots \si_{n-3}) ~\right] \nonumber\\
&& \hspace{6.5cm} ~~ \times  \bra 0| J_1 J_{\si_2} \cdots J_{n} |0
\ket \label{g4}\eeqar where the indices of $T$ and $\tilde{J}$
obey the ordering of $(\si_2 \si_3 \cdots \si_{n-3}\, 1 \, n-1 \,
\tau_{n-2} \, n)$ and those of $J$ obey the ordering of $(1 \si_2
\cdots \si_{n-3} \tau_{n-2} n-1 \, n)$. Since $J$ and $\tilde{J}$
are independent of each other, their expectation values decouple.
With this expression, we can rewrite $\M$ as \beq \M(12\cdots n)
~\sim ~\sum_{\{\si, \tau\}}~\left[ ~ \bra 0| \J_1 \J_2 \cdots \J_n
|0 \ket ~+~ \P(\si_2 \si_3 \cdots \si_{n-3})~\right] \label{g5}
\eeq where the sum is taken over the elements $\{ \si_2, \si_3,
\cdots, \si_{n-3}, \tau_{n-2} \} = \{2,3,\cdots, n-2 \}$ such that
$\si_2 < \si_3 < \cdots < \si_{n-3}$, as considered in
(\ref{lap8}). Note that this amplitude has an analogous form to
the gluon amplitude in (\ref{g2}), especially when we rewrite $\A$
as \beq \A(12\cdots n) \sim \sum_{\si_2 < \si_3 < \cdots < \si_n}
~\left[~\bra A_{1} A_{\si_2} A_{\si_3} \cdots A_{\si_n} \ket ~+ ~
\P(\si_2 \si_3 \cdots \si_n)~\right] \label{g6} \eeq where the
ordering imposition on $\si$ fixes the elements $(\si_2, \cdots,
\si_n)$ to be $(2,\cdots ,n)$.

Since $\A$ and $\M$ have a similar structure, by analogy with the
Yang-Mills (or open string) theory, we can regard a Chan-Paton
factor of $\M$ as $\left[ (T_{\si_2} \tilde{J}_{\si_2}) (T_{\si_3}
\tilde{J}_{\si_3}) \cdots (T_{n} \tilde{J}_{n}) \right]_{perm}$ in
(\ref{g4}) with an understanding of the overall `homogeneous' sum.
This implies that the generators of the local symmetry
(diffeomorphism) for general relativity can be expressed in terms
of $(T_{\si_2} \tilde{J}_{\si_2})$ and so on. One can also find
such an interpretation, noticing the similarity between $A_1 =
t^{a_1} A_{1}^{a_1}$ for $\A(12\cdots n)$ and $\J_1 =(T_{\si_2}
\tilde{J}_{\si_2}) J_1$ for $\M(12\cdots n)$. Apart from the
summations in (\ref{g5}) and (\ref{g6}), the essential difference
between $\A$ and $\M$ lies in the combinatoric structure carried
by the ingredients of the Chan-Paton factor as shown in the
expression (\ref{g4}).

We need to emphasize that an explicit form of $\J_1 \J_2 \cdots
\J_n$ persistent to the expression (\ref{g5}) is not known except
for the MHV amplitude, {\it i.e.}, the scattering amplitude of two
negative helicity gravitons and $n-2$ positive helicity gravitons.
However, we can expect that the expression (\ref{g5}) holds for
generalized amplitudes as well because $\M(12\cdots n)$ does not
depend on any helicity configurations of scattering gravitons. It
is known that all multigluon tree level amplitudes can be
described in terms of the MHV amplitudes by use of the so-called
CSW rule \cite{CSW1}. The rule basically says that the non-MHV
amplitudes can be constructed by a sum of the terms like
$\A_{MHV}(12\cdots k r)~ \frac{1}{{p_{rs}}^2}~ \A_{MHV}(s\,
k+1\cdots n)$ for $1<k<n-1$, where $p_{rs}$ is an internal
momentum transferred between the vertices labelled by $r$ and $s$.
The non-MHV version of $\C(12\cdots n)$ corresponding to this term
would be written as $\<0|J_1 J_2 \cdots J_k |r\>
\frac{1}{{p_{rs}}^2} \<s|J_{k+1} J_{k+2} \cdots J_n |0\>$, where
we denote $|r \>$ for $J_r |0 \>$. This means that any sort of
$\C$ can be described by an expectation value, as we have
expected. The simple form in (\ref{g5}), however, suggests
something more than the CSW rule. It implies a persistent
expression for the graviton amplitudes, regardless their helicity
configurations. In order to understand this well, one may need to
reinterpret the current operator $J$ or $\tilde{J}$ in the manner
that twistor string theory suggests \cite{Witten1}. We shall not
pursue for this interesting subject, rather, in what follows, we
will consider how the Chan-Paton factor arises in the graviton
amplitudes.

The quantity $\J$ in general is composed of $J$ and $\tilde{J}$,
each of which corresponds to an Abelian current of a WZW model
with $\N=4$ supersymmetry. This naturally leads to $\N=8$
supersymmetry for the (composite) current $\J$. One of the crucial
points in Nair's original idea \cite{Nair1} to obtain a
four-dimensional theory out of a usual two-dimensional WZW current
was to realize the fact that the twistor space $\cp^3$ is a
$\cp^1$ bundle over compactified spacetime (or $S^4$). The
non-Abelian current $A_m=t^{a_m}A_{m}^{a_m}$ ($m=1,2,\cdots ,n$)
was obtained by an attachment of a four-dimensional superfield
(which contains information on the $\N=4$ superparticles) to the
current of a WZW model on $\cp^1$, where this $\cp^1$ was
identified with the $\cp^1$ fiber of twistor space. The WZW model
was then naturally extended in the supertwistor space $\cp^{3|4}$,
and the multigluon MHV amplitude could be interpreted as a current
correlator of that extended WZW model. The expression (\ref{g5})
suggests that the multigraviton MHV amplitude can be interpreted
similarly to the multigluon case, except that the correlator of
currents $A$ is replaced by the expectation value of a series of
fixed-ordered composite currents $\J$. This current, say, $\J_1 =
(T_{\si_2}\tilde{J}_{\si_2})J_1$ can be reinterpreted as an analog
of $A_1 =t^{a_1}A_{1}^{a_1}$ such that
$(T_{\si_2}\tilde{J}_{\si_2})$ corresponds to a
(non-supersymmetric) Chan-Paton factor and $J_1$ corresponds to a
WZW current on $\cp^1$ attached with a $\N=8$ superfield. Notice,
however, the Chan-Paton factor of $\J_1$ is not necessarily
$(T_{\si_2}\tilde{J}_{\si_2})$ but any one of the
$(T_{\si_r}\tilde{J}_{\si_r})$ for $r=2,3,\cdots, n-3$. With this
combinatoric structure understood, we may consider this $\J_1$ as
a chiral current of a WZW model on $\cp^1$ whose target space is
$\cp^{3|4}\times \cp^{3|4}$.

Once Grassmann integrals are carried out in the computation of
amplitudes, the four-dimensional superfield that we have mentioned
above can be considered as a scalar field (which is possible by
use of the two-component spinor momenta). This implies that
gravitational fields may be described by scalar fields in a
particular framework of twisor space. It would be nice to realize
that if scalar fields are relevant to gravitational fields, a
direct analogy between Yang-Mills theory and general relativity
can be made because of the following reasons. One can formulate
general relativity as a gauge theory by introducing the following
covariant derivative \beq D_\mu = \d_\mu + e^{m}_{\mu} \d_m +
\Om^{mn}_{\mu} \Si^{mn} \label{g7}\eeq where $e^{m}_{\mu}$,
$\Om^{mn}_{\mu}$ and $\Si^{mn}$ are the frame vector field, the
spin connection and the Lorentz generator, respectively ($\mu,
m=1,2,3,4$). The derivative $\d_m = i p_{m}$ is the translation
operator on the tangent space. In the two-component notation, the
covariant derivative is written as $D_{A \Ad} = (\si^\mu)_{A \Ad}
D_\mu$, where $\si^\mu = (1, -i \si^i)$, with $\si^i$ being the
Pauli matrix. One can incorporate supersymmetry by introducing
spinorial covariant derivatives and spinorial versions of $\d_m$,
while the Lorentz generator remains the same. This leads to the
superspace formulation of supergravity (see, for example,
\cite{vanN1}). If scalar fields are relevant to the gravitational
fields, one can disregard the contributions from $\Om^{mn}_{\mu}
\Si^{mn}$ because, for any scalar filed $\p$, $\Si^{mn} \p =0$ is
satisfied. With this condition assumed, the gravitational data are
solely governed by the frame vector $e_{\mu}^{m}$ whose Chan-Paton
factor can be regarded as $i p_m$.

We may define the metric as a square of $e_\mu = i e_{\mu}^{m}
p_m$, which gives rise to a square of momentum for a Chan-Paton
factor of graviton field. This naturally explains the factors
$T_{\si_2},T_{\si_3}, \cdots T_{\si_{n-3}}$ in (\ref{g1}). Notice
that the way $e_\mu$ couples to each other is nontrivial as we can
see from (\ref{b3}). Further, as we have discussed, we may choose
any one of $T_{\si_2},T_{\si_3}, \cdots T_{\si_{n-3}}$ for the
Chan-Paton factor of a single graviton field. Note that the three
gravitons corresponding to $T_1 = T_{n-1} = T_{n}=1$ are always
irrelevant to such a choice. This is related to the fact that the
KLT-inspired amplitude, as derived from string theory, preserves
$SL(2,{\bf R})$ global symmetry. The graviton corresponding to
$T_{\tau_{n-2}}$ is also irrelevant for fixed $\tau_{n-2}\in
\{2,3,\cdots, n-2\}$. The Chan-Paton factor of a multigraviton
field is then expected to have the factor
$T_{\si_2}T_{\si_{3}}\cdots T_{\si_{n-3}} T_{1}T_{n-1}
T_{\tau_{n-2}} T_{n}$ plus the permutations over $\{\si_2 \si_3
\cdots \si_{n-3}\}$. This gives a plausible interpretation to the
same factor appearing in (\ref{g4}). These discussions are
applicable to the subamplitudes of $\M$ for fixed $\tau_{n-2}$. In
order to reproduce the full KLT-inspired amplitude $\M$, we need
to take the overall `homogeneous' sum. The fact that the metric is
proportional to one of the $T$'s implies that there is no
conformal invariance, which is in accord with general relativity.
The definition of $e_\mu$ in the form of $e_\mu = i e_{\mu}^{m}
p_m$ actually brings a negative sign to each element of
$T_{\si_2}, T_{\si_3}\cdots T_{\si_{n-3}}, T_{\tau_{n-2}}$, which
naturally explains the overall $(-1)^{n+1}$ sign in $\M(12\cdots
n)$. The other overall factor $\ka^{n-2}$ can easily be obtained
by scaling the $J$'s as in the Yang-Mills case.


To summarize, we have presented various expressions for the
multigraviton tree level amplitude (which we have simply called
the graviton amplitude $\M$ in many cases). All expressions are
inspired from the Kawai-Lewellen-Tye (KLT) relation in string
theory and as a result they provide alternatives of the general
formula given by Bern {\it et al}. The similarity between the
gluon amplitude and the graviton amplitude indicates a gauge
theoretic description of $\M$ whose Chan-Paton factor carries a
combinatoric structure in terms of $T_{\si_2},T_{\si_3} \cdots
T_{\si_{n-3}}$. These $T_\si$'s are expressed in terms of a square
of the external momenta for gravitons as shown in (\ref{b3}). The
gauge theoretic interpretation of $\M$ with such a Chan-Paton
factor has been suggested for the maximally helicity violating
(MHV) amplitudes. We have shown that the same interpretation is
possible for the non-MHV amplitudes as well. By analogy with the
Yang-Mills case, we have also seen that the graviton MHV amplitude
is related to a chiral current of a WZW model on $\cp^1$ whose
target space is $\cp^{3|4}\times \cp^{3|4}$ with the above
mentioned combinatoric structure. A relation between the graviton
amplitudes and $\N=8$ supergravity is currently under
investigation.

\vskip .2in\noindent {\bf Acknowledgments} \vskip .06in\noindent
The author would like to thank Professor V.P. Nair for helpful
discussions.


\end{document}